\documentclass[preprint,3p,12pt,review,authoryear]{elsarticle}
\usepackage{natbib}
\usepackage{times,graphics,amssymb,amsmath,epsfig,subfigure,relsize,xcolor}
\usepackage{hyperref}

\journal{New Astronomy}

\begin{document}

\begin{frontmatter}

\title{Broadening of the thermal component of the prompt GRB emission due to rapid temperature evolution}
\author[1,4]{Priya Bharali}
\ead{priya\_phy@gimt-guwahati.ac.in}
\author[2]{Sunder Sahayanathan}
\ead{sunder@barc.gov.in}
\author[3]{Ranjeev Misra}
\author[4]{Kalyanee Boruah}
\address[1]{Girijananda Chowdhury Institute of Management and Technology, Guwahati, India}
\address[2]{Bhabha Atomic Research Centre, Mumbai, India}
\address[3]{IUCAA, Pune, India}
\address[4]{Department of Physics, Gauhati University, Guwahati, India}

\begin{abstract}
The observations of the prompt emission of gamma ray bursts (GRB) by GLAST Burst Monitor (GBM), on 
board \emph{Fermi} Gamma-ray Space Telescope, suggest the
presence of a significant thermal spectral component, whose origin is
not well understood. Recently, it has been shown that  for long duration
GRBs, the spectral width as defined as the logarithm of the ratio of the
energies at which the spectrum falls to half its peak value,
 lie in  the range of 0.84-1.3 with a median value of 1.07. Thus,
 while most of the GRB spectra are found to be too narrow to be
explained by synchrotron emission from an electron distribution, they are also significantly broader 
than a blackbody spectrum whose width should be 
0.54.  Here, we consider the
possibility that an intrinsic thermal spectrum from a  fire-ball like model,
may be observed to be broadened if the system undergoes a rapid temperature
evolution. We construct a toy-model to show that for bursts with
durations in the range 5-70 seconds, the widths of their 1 second time-averaged spectra can be at 
the most $\lesssim 0.557$. Thus, while rapid temperature variation
can broaden the  detected spectral shape, the observed median value of $\sim 1.07$
requires that there must be significant sub-photospheric emission and/or an anisotropic
explosion to explain the broadening for most GRB spectra.
\end{abstract}


\end{frontmatter}

\section{Introduction}
Gamma ray bursts (GRB) are the most luminous, transient phenomena happening in the universe with luminosities
of the order of $10^{51}-10^{52}$ erg/s \citep{Piran04}. The prompt phase of the GRB corresponds to the
initial episodic event which lasts typically for a duration ranging from a fraction of a second 
to few tens of seconds. This is often followed by a fading emission, long after the initial 
burst decayed, termed as ``after glow''. Optical
study of these after glow confirmed the cosmological origin of GRBs \citep{Costa97,Paradijs97}. The distribution
of the burst duration of GRBs is bimodal with a minima falling at $\sim 2$ s suggesting the 
GRB may plausibly arise from two different process \citep{Kouveliotou93}. Accordingly they were classified into two types, 
namely short bursts with duration $<2$ s and long bursts with duration $>2$ s. Further, the spectra of
short bursts are typically harder than the long ones supporting different origin of these two 
classes \citep{Bhat16,Kouveliotou96,Dezalay96}. The progenitor of the GRBs are not well understood; however,
observational and theoretical advancements suggests short GRBs to be associated with the mergers of compact objects, e.g. 
neutron star-neutron star merger or neutron star-black hole merger\citep{Eichler89,Paczynski91,Rosswog13}, 
and the long ones are associated with the
collapse of a massive star onto a black hole\citep{Woosley93,MacFadyen99}.

Initial attempts to understand the GRB emission were focussed on thermal nature originating 
from a catastrophic event involving collapse of a massive star or merging of two compact
objects \citep{Paczynski86,Goodman86}. The huge energy released within a small volume implies, the  
optical depth of the initial medium to be very high and the interaction between the energetic photons 
and particles will cause the medium to expand at relativistic speed, commonly referred as
``fire-ball''. However, after sufficient
expansion, the fire-ball approaches the photospheric stage where further decrease in density transits the 
matter to optically thin regime and the trapped thermal photons are released \citep{Goodman86,Burgess15}. 
On the contrary, the time-averaged spectra of GRBs, observed by \emph{BATSE} on board
\emph{CGRO}, are found to be non-thermal and are well explained by a broken power-law function with smooth
transition at the break frequency (Band function) \citep{Band93}.
To understand this non-thermal emission, internal shock models were proposed where particles are accelerated at a shock front 
initiated by the collision between shells of matter expelled by the initial catastrophic 
event \citep{Kobayashi97,Panaitescu98,Daigne98}. 
The GRB emission is then modelled as the synchrotron emission from the relativistic electron
population accelerated at these shock fronts. Alternatively, \cite{Blinnikov99} showed that
a non-thermal spectrum can also be imitated by the time integration of blackbody emission 
arising from an evolving GRB shells.

The observed photon spectral indices of many GRBs; however, confront the synchrotron emission 
interpretation as the indices are steeper than the one allowed by this model\citep{Crider97,Preece98,Ghirlanda03}. 
In addition, the low energy conversion efficiency of internal shock models posed severe 
drawback \citep{Ryde04,Pe'er16}. These 
discrepancies forced the addition of a thermal component in the GRB spectra again, along with the 
non-thermal one \citep{Guiriec10,Guiriec13,Zhang11,Burgess14,Axelsson12}.
After the advent of \emph{Fermi} and \emph{Swift}, satellite based experiments operating at gamma ray 
and X-ray energies, the presence of thermal component in GRB spectrum became more evident\citep{Basak14,Rao14}.

Recently, \cite{Axelsson15} performed an elaborative spectral study of long and short bursts observed by Fermi/GBM and 
CGRO/BATSE and compared their spectral widths. The full width half maximum ($W$) of the $EF_E$ representation of the spectra, 
with $E$ being the photon energy and $F_E$ the specific flux, was calculated during the flux maximum of each bursts by fitting the 
observed fluxes using a Band function. Interestingly, the width distribution peaked at $\sim 1.07$ for long GRBs and $\sim 0.86$ 
for short GRBs, which is much broader than the Planck function ($W \approx 0.54$) but narrower than the synchrotron spectrum due
to a Maxwellian distribution of electrons ($W = 1.4$) or a power-law electron distribution with index $-2$ ($W=1.6$).
A similar study was also carried out by \cite{Yu15} who studied the spectral curvature at the peak of the GRB spectrum for 1113 bursts 
detected by the {\emph{Fermi}} GBM experiment. Again, they concluded that most of the bursts are inconsistent with synchrotron 
emission models or a single temperature blackbody emission. In case of short bursts, a detailed study of spectral broadening
was performed by \cite{Begue14} using approximate analytical solution for the relativistic hydrodynamic equations.

In the present work, we study the broadening of the photospheric thermal emission due to relativistic effects under the fire-ball model
of GRBs. We consider a scenario where the temperature of the fire-ball decreases rapidly with increase in radius and this causes the high
latitude emission to be relatively hotter than the on-axis emission for a distant observer. In addition, the time integrated
spectrum will cause further broadening due to the evolution of the fire-ball within the integration time. Particularly,
we investigate the maximum attainable width of this multi temperature blackbody emission within the burst timescale typical for
long GRBs. In the next section, we describe the model and the spectral properties. In \S 3, we study the maximum attainable 
width under the present model for the case of long GRBs and summarize the work in \S4.

\section{Spectral Evolution of an Expanding Fire-ball}
\label{sec:fb}
We consider the thermal emission from the photosphere of GRB to be associated with the release of radiation trapped in an initial optically thick and
relativistically expanding ball of plasma (Fire-ball). We also assume the expansion to be associated with a rapid fall in temperature
described by
\begin{align}\label{eq:tempvariation}
	T(r)= T_0 \left(\frac{r_0}{r}\right)^{\psi}
\end{align}
Here, $T$ is the temperature when the fire-ball radius is $r$, $T_0$ corresponds to the temperature at radius $r_0$
and $\psi$ is the temperature index describing the cooling. 
The burst is assumed to be in its coasting phase and hence 
the expansion velocity will be approximately constant till the internal shock or other dissipative events occur \citep{Vedrenne09}. Further, if the expansion is adiabatic, conservation 
of entropy limits the value of $\psi$ to be $2/3$ during the coasting phase of the burst \citep{Piran93,Pe'er15}.

\begin{figure}
\begin{center}
\includegraphics[width=0.5\textwidth,angle = -90] {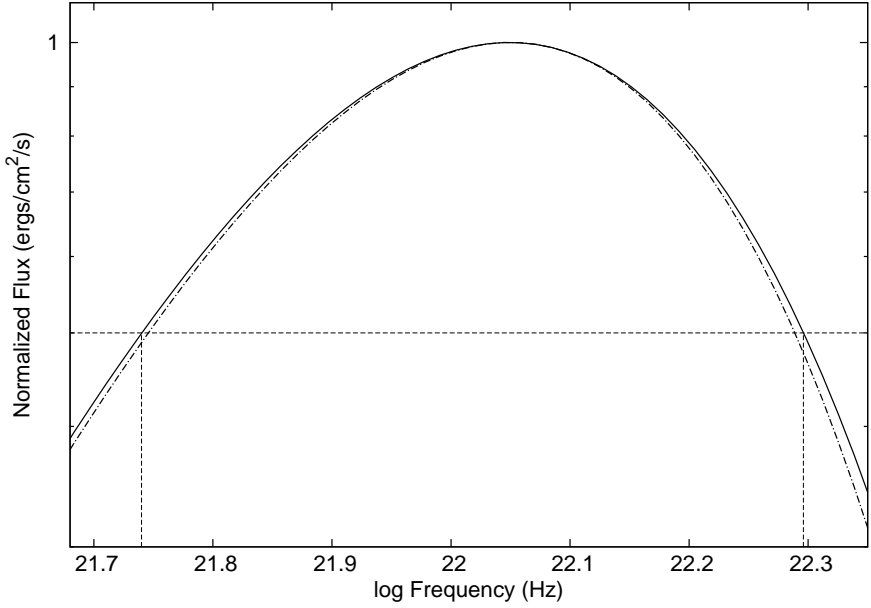}
\caption{The instantaneous spectrum corresponding to $\Gamma$ = 500 and $\psi$ = 2/3 
for $T_0$ = 10 keV. The dashed line indicates the spectral width W measured at FWHM.
The dot-dashed line represents equivalent blackbody spectrum. }\label{fig:instant_spec}  
\end{center}
\end{figure}

The relativistic expansion of the fire-ball and the light travel time
effect will cause the emission from a higher latitude to be hotter than the on axis emission for a distant observer. The flux
at frequency $\nu$ received by the observer located at a distance $D$ will then be a modified blackbody spectrum given by (Appendix \ref{appendix:instant_spec})
\begin{align}
	f_\nu(r_0) = \frac{4\pi h}{c^2} \frac{r_0^2}{D^2}(1-\beta)^2 \nu^3 \int\limits^1_\beta  
	\frac{\mu\, d\mu}{\left\{ \textrm{exp}\left[\frac{h\nu}{\Gamma k T_0(1+\beta\mu)}\left(\frac{1-\beta}{1-\beta\mu}\right)^\psi \right]-1\right\} }
	\frac{(\mu-\beta)}{(1-\beta\mu)^3}
\end{align}
where, $\beta$ is the expansion velocity in units of the speed of light $c$ and $\Gamma=(1-\beta^2)^{-1/2}$ is the corresponding Lorentz factor,
$r_0$ is the on axis radius of the fire-ball measured by the observer, $\mu$ is the cosine of 
the latitude, $h$ is the Planck constant and $k$ is the Boltzmann constant. If we consider the 
photosphere to be spherical, then the relativistic beaming effects will cause the off-axis emission 
received by the observer to be limited within an angle $1/\Gamma$ subtended at the centre
of the fire-ball. When the comoving plasma density varies as $r^{-2}$ and for an energy independent 
photon scattering cross section, the photosphere can be significantly different from a simple 
sphere \citep{Abramowicz91}. However, within the opening angle $1/\Gamma$ along the line of sight of the observer, 
the surface of the photosphere can still be approximated as spherical \citep{Pe'er08}. For simplicity,
the photosphere 
emission beyond this angle is not considered in the present work. 
In Fig. \ref{fig:instant_spec}, we show the instantaneous spectrum (normalized) 
for $\Gamma=500$, $\psi=2/3$ and $T_0=10$ keV along 
with equivalent blackbody spectrum. 
For $\Gamma\gg1$, as in the case of GRBs, the emission cone will be narrower and the instantaneous spectrum observed 
will drift towards a simple blackbody, unless $\psi$ is large enough to create significant off-axis temperature variation. 

\begin{figure}
\begin{center}
\includegraphics[width=0.5\textwidth,angle = -90] {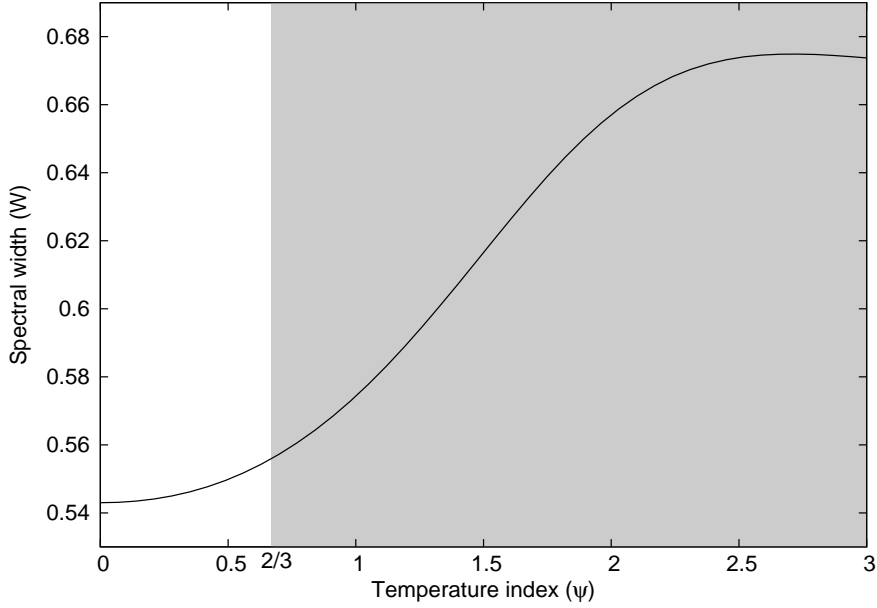}
\caption{Variation of the instantaneous spectral width W with respect to $\psi$}\label{fig:inst_w_psi}  
\end{center}
\end{figure}

Following \cite{Axelsson15}, if we define the width of the resultant spectrum as (Fig. \ref{fig:instant_spec}) 
\begin{align}
	W=\textrm{log}\left(\frac{\nu_2}{\nu_1}\right) 
\end{align}
where, $\nu_1$ and $\nu_2$ are the photon frequencies at the full width at half maximum (FWHM) of the 
$\nu f_{\nu}$ (unit: $ergs/cm^2/s$) spectrum, then for $\psi>0$, $W$ will be larger than the one expected from a  
simple blackbody spectrum ($W \approx 0.54$). In addition, for $\Gamma \gg 1$, $W$
will depend mainly on $\psi$ and in Fig. \ref{fig:inst_w_psi} we show its variation with respect to the latter. 
As $\psi$ increases, $W$ increases from $\sim 0.54$ to a maximum of $\approx 0.675$ corresponding to 
$\psi \approx 2.7$. Beyond this, the temperature gradient becomes too large such that the 
on-axis emission fall below the FWHM of the hotter high latitude emission, causing $W$ to decrease with
further increase in $\psi$. Under the adiabatic limit ($\psi=2/3$), the maximum value of W that can
be attained is $\approx 0.556$ and beyond this range is shown as the shaded region in Fig. \ref{fig:inst_w_psi}.

\begin{figure}
\begin{center}
\includegraphics[width=0.5\textwidth,angle = -90] {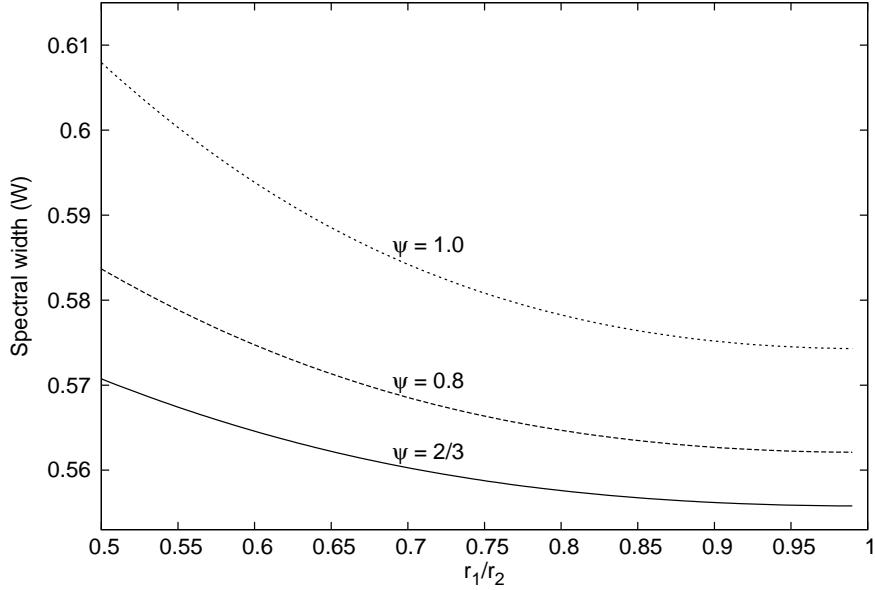}
\caption{Variation of the time-averaged spectral width W with respect to the ratio of initial and 
final radius $r_1/r_2$ corresponding to the time segment, for $\psi=$ $2/3$, $0.8$ and $1.0$.}\label{fig:wr1r2}  
\end{center}
\end{figure}

A time-averaged spectrum will involve the temporal evolution of the fire-ball within the duration
and this will further broaden the spectrum. Considering the fire-ball expands from a radius $r_1$ to $r_2$
during an interval, the time-averaged spectrum will then be
\begin{align} \label{eq:r_ave}
	F_{1,2}(\nu)=\frac{\int\limits_{r_1/r_2}^1 \bar{f}_\nu(x) dx}{1-r_1/r_2}
\end{align}
where, $\bar{f}_\nu(x)=f_\nu(xr_2)$ and the corresponding temperature $T_x=T_2 \; x^\psi$ 
with $T_2$ the temperature of the fire-ball at radius $r_2$.
In Fig. \ref{fig:wr1r2}, we show the dependence of $W$ on the ratio $r_1/r_2$ for 
the case $\Gamma\gg1$ and $\psi=2/3$, $0.8$ and $1.0$.
If the time-averaged spectrum is obtained for a time step $\Delta$
consecutively over the entire burst, then the radius of the fire-ball after $n\Delta$ duration
can be expressed in terms of the initial radius of the burst ($r_0$) as
\begin{align}
	r_n &= n\beta c \Delta \Gamma^2+r_0 \nonumber \quad 
\end{align}
The ratio of the radii falling on the beginning and the end of $\Delta$ will then be
\begin{align}\label{eq:rratio}
	\frac{r_n}{r_{n+1}} &= 1-\frac{1}{n+1+\xi}
\end{align}
where, $\xi = \frac{r_0}{\beta c \Delta \Gamma^2}$. Since the ratio of the radii continuously evolves during a burst, the width of the time-averaged
spectrum will additionally depend on the initial burst radius $r_0$ and the time segment $n$, along with 
$\Gamma$ and $\psi$. From equation (\ref{eq:rratio}), the minimum value of the ratio of 
radii attained will be $0.5$ corresponding to $\xi = 0$ and $n=1$, which approaches to $1$ as $n$ increases.
Hence, $W$ will eventually 
attains a constant value with negligible change soon after the explosion.
The light curve of the burst can be obtained by integrating equation (\ref{eq:r_ave}) over the frequency range of interest.
The duration of the burst ($\tau_{90}$) can then be obtained by clipping the light curve 
at 5\% of the total flux in the beginning and the 
end of the burst. 

\section{Spectral width of Long Bursts}
\begin{figure}
\begin{center}
\includegraphics[width=0.5\textwidth] {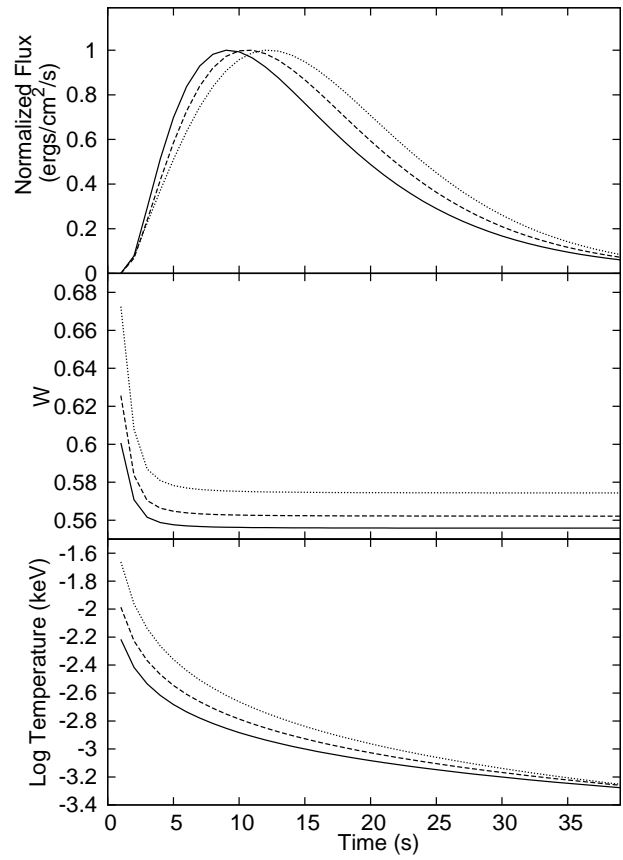}
\caption{The evolution of the normalized flux (top), time-averaged spectral width W (middle) 
and temperature(bottom) corresponding to $\Gamma = 500$ and $T_0 = 10 keV$ for 
$\psi=$ $2/3$ (bold), $0.8$ (dashed) and $1.0$ (dotted)}\label{fig:longgrb}  
\end{center}
\end{figure}

To study the maximum attainable width of long GRBs under the expanding fire-ball scenario, 
we obtain one second time-averaged light curve integrated over the energies 8 keV to 40 MeV, 
consistent with the observations \citep{Axelsson15}. 
In Fig. \ref{fig:longgrb}, we show the evolution of the normalized flux, time-averaged spectral
width $W$ and temperature corresponding to $\Gamma=500$ and $T_0=10$ keV.
 The parameters $\psi$ and $\xi$ are chosen to be 
($2/3$, $1.5\times 10^{-5}$), ($0.8$, $1.85\times 10^{-4}$) and ($1.0$, $2.18\times 10^{-3}$)
such that
$\tau_{90}\approx 30$ s, the typical duration of long GRBs. We find that the choice of
$\psi$ significantly varies the temporal profile, with
the burst peaking earlier for smaller $\psi$ values. For a given $\psi$, the burst
peak time ($t_{peak}$) can be elongated by increasing $\xi$; nevertheless, this will also increase the $\tau_{90}$ from
the desired value.

The spectral width $W$ can be as large as $\sim 0.6$, during the initial phase of the burst; however,  
it rapidly decreases to a nearly constant value due to increase in ratio of radii $\frac{r_n}{r_{n+1}}$. 
Maintaining a constant $\tau_{90}$, the width of the spectrum during the flux peak ($W_p$) increases with 
$\psi$ and, in principle, can be adjusted to obtain the desired value $\approx 1.07$. For example, choosing 
$\xi = 0.31$, $\Gamma = 600$ and $\psi=2.14$, we obtain $W_p\approx1$ while $\tau_{90}\approx30$ s; however, such choices will 
cause the burst to peak much earlier ($t_{peak}<1$s), inconsistent with the observations. Reduction of $\psi$ 
{ to the adiabatic limiting value $2/3$} reduces $W_p$ 
and this should be associated with a decrease in initial radius $r_0$ in addition, 
to maintain $\tau_{90}$ at the desired value.

\begin{figure*}
\begin{center}
\includegraphics[width=0.5\textwidth,angle = -90] {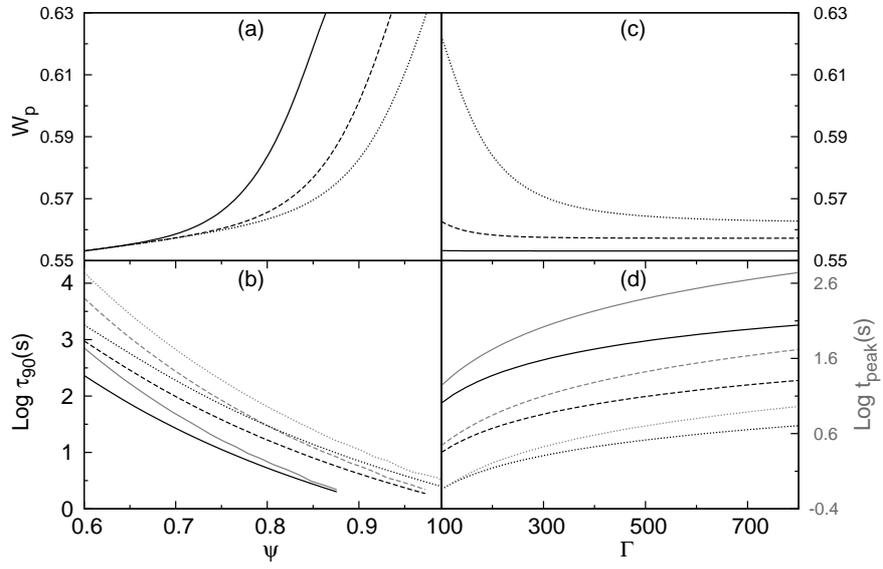}
\caption{Left (a) and (b): Variation of the spectral width during the burst peak, $W_p$ (a), and $\tau_{90}$ (b) 
with respect to $\psi$ for $\Gamma=$ $200$ (solid), $500$ (dashed) and $800$ (dotted). 
Right (c) and (d): Variation of the spectral width during the burst 
peak, $W_p$ (c), and $\tau_{90}$ (d) with respect to $\Gamma$ for $\psi=$ $0.6$ (solid), $0.7$ (dashed) and 
$0.8$ (dotted). Gray lines in (b) and (d) indicate the burst peak time $t_{peak}$.}\label{fig:longgrb1}  
\end{center}
\end{figure*}

For a comparative study between the fire-ball parameters and the derived quantities of the burst, we study the 
variation of $W_p$ and $\tau_{90}$ 
with respect to $\psi$ and $\Gamma$. { In Fig. \ref{fig:longgrb1}a and \ref{fig:longgrb1}b, we 
show the variation of $W_p$, $\tau_{90}$ and $t_{peak}$ with $\psi$
for $\Gamma =$ $200$ (solid), $500$(dashed) and $800$(dotted). The value of $\xi$ is chosen to be $10^{-4}$ and $T_0$ is fixed at 
$10$ keV. For a given $\Gamma$, $W_p$ increases with $\psi$ and sharply beyond $\psi\sim 0.8$. However, 
this is also associated with a significant decline in $\tau_{90}$ and $t_{peak}$ (gray lines). 
In Fig. \ref{fig:longgrb1}c and \ref{fig:longgrb1}d, we  again show the variation of $W_p$, $\tau_{90}$ and $t_{peak}$
with respect to $\Gamma$ for $\psi=$ $0.6$ (solid), $0.7$ (dashed) and $0.8$ (dotted)}. Here, larger 
$\Gamma$ is associated with delayed $t_{peak}$ and hence narrower $W_p$ (see Fig. \ref{fig:longgrb}). 
Hence, requirement of a broader $W_p$ demands a larger $\psi$ 
and a smaller $\Gamma$ which on the other hand shortens the burst duration { as well as deviate
largely from the adiabatic limit}. 
This enforces a limit on maximum attainable width under the assumed fire-ball scenario and 
significantly hampers to achieve the width ($1.07$) demanded by the observations.
{ Through the present study, we found that with an optimal choice of parameters and 
$\psi=2/3$, 
maintaining $\tau_{90}$ and $t_{peak}$ at a reasonable values corresponding to long GRBs,
one can only attain $W_p\lesssim 0.557$ (e.g. $W_p\approx 0.557$ with $\tau_{90}\approx 20$ s and $t_{peak}\approx 5$ s
for $\Gamma=100$ and $\xi=1.12\times10^{-4}$ and $\psi=2/3$)}.

\section{Summary \& Discussion} 

The FWHM spectral width of the Band spectrum used to fit the time resolved spectrum of long GRBs is observed to be 
$\approx 1.07$. This is larger than the width of the simple blackbody spectrum defined
by a Planck's function and smaller than the synchrotron spectrum from a Maxwellian/power-law distribution 
of electrons \citep{Axelsson15}. 
To understand the maximum attainable width under a simple fire-ball interpretation of GRB,
we consider a scenario where the GRB emission 
to be a multi temperature blackbody distribution arising from a rapidly 
cooling and relativistically expanding fire-ball of hot thermal plasma. The dynamics of the fire-ball will cause 
the instantaneous spectrum to be broader than the simple blackbody spectrum with the emission from the higher latitudes 
hotter than the on-axis emission due to light travel time and relativistic effects. 
In addition, the time averaging of the emission will incorporate considerable evolution of the fire-ball and
this will further broaden the emission. 
However, under this scenario, we are only able to obtain a maximum spectral width of $\approx 0.557$, 
while maintaining the burst duration to be consistent with the long GRBs. This is marginally broader 
than the simple blackbody spectrum but significantly smaller than the observed values.

In this work, we highlight the inadequacy of a simple fire-ball model to explain the observed 
large spectral width of the GRB prompt emission. 
A plausible explanation for this may include the emission from the sub-photospheric region.
The last
scattering position of the trapped photons may not be necessarily associated with the photosphere, rather can 
emerge from regions beneath it \citep{Beloborodov10,Pe'er08}. Hence, the emission could be from 
an extended volume instead of a simple surface which can significantly broaden the observed spectrum.  Alternatively,
the expansion velocity of the fire-ball may not be isotropic, instead depend on the opening angle \citep{Lundman13}. 
Such stratification of flow velocity can include emission from higher latitudes (i.e beyond the opening
angle $1/\Gamma$ of an uniformly expanding fire-ball) which can further broaden the observed spectrum. 

\def\aj{AJ}%
\def\actaa{Acta Astron.}%
\def\araa{ARA\&A}%
\def\apj{ApJ}%
\def\apjl{ApJ}%
\def\apjs{ApJS}%
\def\ao{Appl.~Opt.}%
\def\apss{Ap\&SS}%
\def\aap{A\&A}% 
\def\aapr{A\&A~Rev.}%
\def\aaps{A\&AS}%
\def\azh{AZh}%
\def\baas{BAAS}%
\def\bac{Bull. astr. Inst. Czechosl.}%
\def\caa{Chinese Astron. Astrophys.}%
\def\cjaa{Chinese J. Astron. Astrophys.}%
\def\icarus{Icarus}%
\def\jcap{J. Cosmology Astropart. Phys.}%
\def\jrasc{JRASC}%
\def\mnras{MNRAS}%
\def\memras{MmRAS}%
\def\na{New A}%
\def\nar{New A Rev.}%
\def\pasa{PASA}%
\def\pra{Phys.~Rev.~A}%
\def\prb{Phys.~Rev.~B}%
\def\prc{Phys.~Rev.~C}%
\def\prd{Phys.~Rev.~D}%
\def\pre{Phys.~Rev.~E}%
\def\prl{Phys.~Rev.~Lett.}%
\def\pasp{PASP}%
\def\pasj{PASJ}%
\def\qjras{QJRAS}%2215.bib
\def\rmxaa{Rev. Mexicana Astron. Astrofis.}%
\def\skytel{S\&T}%
\def\solphys{Sol.~Phys.}%
\def\sovast{Soviet~Ast.}%
\def\ssr{Space~Sci.~Rev.}%
\def\zap{ZAp}%
\def\nat{Nature}%
\def\iaucirc{IAU~Circ.}%
\def\aplett{Astrophys.~Lett.}%
\def\apspr{Astrophys.~Space~Phys.~Res.}%
\def\bain{Bull.~Astron.~Inst.~Netherlands}%
\def\fcp{Fund.~Cosmic~Phys.}%
\def\gca{Geochim.~Cosmochim.~Acta}%
\def\grl{Geophys.~Res.~Lett.}%
\def\jcp{J.~Chem.~Phys.}%
\def\jgr{J.~Geophys.~Res.}%
\def\jqsrt{J.~Quant.~Spec.~Radiat.~Transf.}%
\def\memsai{Mem.~Soc.~Astron.~Italiana}%
\def\nphysa{Nucl.~Phys.~A}%
\def\physrep{Phys.~Rep.}%
\def\physscr{Phys.~Scr}%
\def\planss{Planet.~Space~Sci.}%
\def\procspie{Proc.~SPIED}%
\let\astap=\aap
\let\apjlett=\apjl
\let\apjsupp=\apjs
\let\applopt=\ao
%
%\bibliographystyle{elsarticle-harv} % (uses file "plain.bst")
%\bibliography{ref} % expects file "myrefs.bib"

\begin{thebibliography}{}
\bibitem[\protect\citeauthoryear{Abramowicz, Novikov, \& Paczynski}{1991}]{Abramowicz91} Abramowicz M.~A., Novikov I.~D., Paczynski B., 1991, ApJ, 369, 175
\bibitem[\protect\citeauthoryear{Axelsson et al.}{2012}]{Axelsson12} Axelsson M., et al., 2012, ApJ, 757, L31 
\bibitem[\protect\citeauthoryear{Axelsson \& Borgonovo}{2015}]{Axelsson15} Axelsson M., Borgonovo L., 2015, MNRAS, 447, 3150 
\bibitem[\protect\citeauthoryear{Band et al.}{1993}]{Band93} Band D., et al., 1993, ApJ, 413, 281 
\bibitem[\protect\citeauthoryear{Basak \& Rao}{2014}]{Basak14} Basak R., Rao A.~R., 2014, MNRAS, 442, 419 
\bibitem[\protect\citeauthoryear{B{\'e}gu{\'e} \& Vereshchagin}{2014}]{Begue14} B{\'e}gu{\'e} D., Vereshchagin G.~V., 2014, MNRAS, 439, 924
\bibitem[\protect\citeauthoryear{Bhat et al.}{2016}]{Bhat16} Bhat P. N., et al., 2016, ApJS, 223, 28 
\bibitem[\protect\citeauthoryear{Beloborodov}{2010}]{Beloborodov10} Beloborodov A.~M., 2010, MNRAS, 407, 1033
\bibitem[\protect\citeauthoryear{Blinnikov, Kozyreva, \& Panchenko}{1999}]{Blinnikov99} Blinnikov S.~I., Kozyreva A.~V., Panchenko I.~E., 1999, Astronomy Reports, 43, 739
\bibitem[\protect\citeauthoryear{Burgess et al.}{2014}]{Burgess14} Burgess J.~M., et al., 2014, ApJ, 784, L43
\bibitem[\protect\citeauthoryear{Burgess \& Ryde}{2015}]{Burgess15} Burgess J.~M., Ryde F., 2015, MNRAS, 447, 3087 
\bibitem[\protect\citeauthoryear{Costa et al.}{1997}]{Costa97} Costa E., et al., 1997, Nature, 387, 783 
\bibitem[\protect\citeauthoryear{Crider et al.}{1997}]{Crider97} Crider A., et al., 1997, ApJ, 479, L39 
\bibitem[\protect\citeauthoryear{Daigne \& Mochkovitch}{1998}]{Daigne98} Daigne F., Mochkovitch R., 1998, MNRAS, 296, 275 
\bibitem[\protect\citeauthoryear{Dezalay et al.}{1996}]{Dezalay96} Dezalay J.~P., et al., 1996, ApJ, 471, L27 
\bibitem[\protect\citeauthoryear{Eichler et al.}{1989}]{Eichler89} Eichler D., Livio M., Piran T., Schramm D.~N., 1989, Nature, 340, 126 
\bibitem[\protect\citeauthoryear{Ghirlanda, Celotti, \& Ghisellini}{2003}]{Ghirlanda03} Ghirlanda G., Celotti A., Ghisellini G., 2003, A\&A, 406, 879 
\bibitem[\protect\citeauthoryear{Goodman}{1986}]{Goodman86} Goodman J., 1986, ApJ, 308, L47
\bibitem[\protect\citeauthoryear{Guiriec \& Fermi/GBM Collaboration}{2010}]{Guiriec10} Guiriec S., Fermi/GBM Collaboration, 2010, Bulletin of the American Astronomical Society, 42, 654 
\bibitem[\protect\citeauthoryear{Guiriec et al.}{2013}]{Guiriec13} Guiriec S., et al., 2013, ApJ, 770, 32 
\bibitem[\protect\citeauthoryear{Kobayashi, Piran, \& Sari}{1997}]{Kobayashi97} Kobayashi S., Piran T., Sari R., 1997, ApJ, 490, 92 
\bibitem[\protect\citeauthoryear{Kouveliotou et al.}{1993}]{Kouveliotou93} Kouveliotou C., et al., 1993, ApJ, 413, L101 
\bibitem[\protect\citeauthoryear{Kouveliotou et al.}{1996}]{Kouveliotou96} Kouveliotou C., et al., 1996, AIPC, 384, 42
\bibitem[\protect\citeauthoryear{Lundman, Pe'er, \& Ryde}{2013}]{Lundman13} Lundman C., Pe'er A., Ryde F., 2013, MNRAS, 428, 2430
\bibitem[\protect\citeauthoryear{MacFadyen \& Woosley}{1999}]{MacFadyen99} MacFadyen A.~I., Woosley S.~E., 1999, ApJ, 524, 262 
\bibitem[\protect\citeauthoryear{Paczynski}{1986}]{Paczynski86} Paczynski B., 1986, ApJ, 308, L43
\bibitem[\protect\citeauthoryear{Paczynski}{1991}]{Paczynski91} Paczynski B., 1991, A\&A, 41, 257
\bibitem[\protect\citeauthoryear{Panaitescu \& Meszaros}{1998}]{Panaitescu98} Panaitescu A., Meszaros P., 1998, ApJ, 526, 707 
\bibitem[\protect\citeauthoryear{Pe'er}{2008}]{Pe'er08} Pe'er A., 2008, ApJ, 682, 463
\bibitem[\protect\citeauthoryear{Pe'er}{2015}]{Pe'er15} Pe'er A., 2015, AdAst, 2015, 907321
\bibitem[\protect\citeauthoryear{Pe'er \& Ryde}{2016}]{Pe'er16} Pe'er A., Ryde, F., 2016, arXiv:1603.05058 
\bibitem[\protect\citeauthoryear{Piran, Shemi, \& Narayan}{1993}]{Piran93} Piran T., Shemi A., Narayan R., 1993, MNRAS, 263, 861 
\bibitem[\protect\citeauthoryear{Piran}{2004}]{Piran04} Piran T., 2004, RvMP, 76, 1143
\bibitem[\protect\citeauthoryear{Preece et al.}{1998}]{Preece98} Preece R.~D., et al., 1998, ApJ, 506, L23 
\bibitem[\protect\citeauthoryear{Rao et al.}{2014}]{Rao14} Rao A. R., et al., 2014, RAA, 14, 35-46 
\bibitem[\protect\citeauthoryear{Rosswog, Piran, \& Nakar}{2013}]{Rosswog13} Rosswog S., Piran T., Nakar E., 2013, MNRAS, 430, 2585
\bibitem[\protect\citeauthoryear{Ryde}{2004}]{Ryde04} Ryde F., 2004, ApJ, 614, 827
\bibitem[\protect\citeauthoryear{van Paradijs et al.}{1997}]{Paradijs97} van Paradijs J., et al., 1997, Nature, 386, 686 
\bibitem[\protect\citeauthoryear{Vedrenne \& Atteia}{2009}]{Vedrenne09} Vedrenne G., Atteia J.-L., 2009, Gamma-Ray Bursts: The Brightest Explosions in the Universe, Springer Praxis Books. Springer, Berlin, Heidelberg  
\bibitem[\protect\citeauthoryear{Woosley}{1993}]{Woosley93} Woosley S.~E., 1993, ApJ, 405, 273 
\bibitem[\protect\citeauthoryear{Yu et al.}{2015}]{Yu15} Yu H.-F., et al.,  2015, A\&A, 583, A129 
\bibitem[\protect\citeauthoryear{Zhang \& Yan}{2011}]{Zhang11} Zhang B., Yan H., 2011, ApJ, 726, 90 
\end{thebibliography}

%\label{lastpage}
\appendix
\section{Instantaneous thermal spectrum from the expanding fire-ball}\label{appendix:instant_spec}
\begin{figure}
\begin{center}
\includegraphics[width=0.5\textwidth] {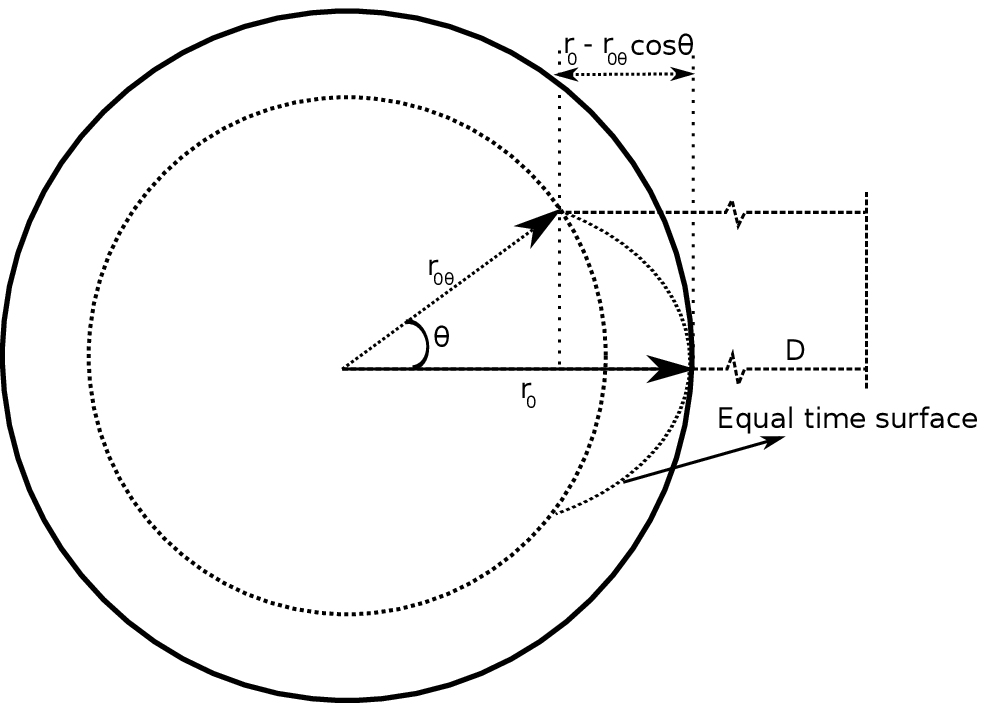}
\caption{Schematic representation indicating the light travel time effects on the instantaneous
spectrum seen by a distant observer.}  
\label{fig:appendix}
\end{center}
\end{figure}

Let $r_0$ be the instantaneous radius of the fire-ball. For an observer located at a distance D,
the emission from $r_0$ will coincide the emission from higher latitude ($\theta$) at an earlier radius 
$r_{0\theta}$ due to light travel time effects (Fig. \ref{fig:appendix}). If the expansion
velocity is $\beta c$ ($c$ being the speed of light), then
\begin{align}\label{eq:radiusrel}
	r_{0\theta}= r_0\left[\frac{(1-\beta)}{(1-\beta cos\theta)}\right]
\end{align}
The net flux at frequency $\nu$ emitted from the surface of the fire-ball will then be
\begin{align}
F_\nu = \frac{2\pi r_0^2\Gamma^3}{D^2}(1-\beta)^2\int\limits_\beta^1 I_{\nu'}\left(\frac{1+\beta\mu}{1-\beta\mu}\right)^3 (\mu-\beta)\mu\, d\mu
\end{align}
Here, $\mu=cos\theta$, $I_{\nu'}$ is the specific intensity measured in the rest frame of the ejecta and $\nu'=\nu[\Gamma(1+\beta\mu)]^{-1}$.
For thermal emission, $I_{\nu'}$ can be replaced by the Planck function and we get
\begin{align}\label{eq:flux1}
	F_\nu = \frac{4\pi h}{c^2}\frac{r_0^2}{D^2}(1-\beta)^2 \nu^3 \int\limits_\beta^1 
	\frac{\mu\,d\mu}{\left\{ \textrm{exp}\left[\frac{h\nu}{\Gamma k T_{0\theta}(1+\beta\mu)}\right]-1 \right\}}\frac{(\mu-\beta)}{(1-\beta\mu)^3}
\end{align}
where, $T_{0\theta}$ is the temperature corresponding to radius $r_{0\theta}$, $h$ is the Planck 
constant and $k$ is the Boltzmann constant. Using equation (\ref{eq:tempvariation}) and (\ref{eq:radiusrel}),
$T_{0\theta}$ can be expressed as
\begin{align}
	T_{0\theta}=T_0\left(\frac{1-\beta\mu}{1-\beta}\right)^\psi
\end{align}
and equation (\ref{eq:flux1}) can be written as
\begin{align}
	f_\nu(r_0) = \frac{4\pi h}{c^2} \frac{r_0^2}{D^2}(1-\beta)^2 \nu^3 \int\limits^1_\beta  
	\frac{\mu\, d\mu}{\left\{ \textrm{exp}\left[\frac{h\nu}{\Gamma k T_0(1+\beta\mu)}\left(\frac{1-\beta}{1-\beta\mu}\right)^\psi \right]-1\right\} }
	\frac{(\mu-\beta)}{(1-\beta\mu)^3}
\end{align}

\end{document}